\documentclass[letterpaper]{JHEP}
\usepackage{epsfig}

\usepackage{amssymb}
\newcommand{\R}{\ensuremath{\mathbb R}}

\newcommand{\Z}{\ensuremath{\mathbb Z}}
\newcommand{\C}{\ensuremath{\mathbb C}}

\newcommand{\CP}[1]{\ensuremath{{\C}{\rm P}^{#1}}}
\newcommand{\RP}[1]{\ensuremath{{\R}{\rm P}^{#1}}}

\newcommand{\ket}[1]{|#1\rangle}
\newcommand{\bra}[1]{\langle#1|}
\newcommand{\be}{\begin{equation}}
\newcommand{\ee}{\end{equation}}
\newcommand{\bea}{\begin{eqnarray}}
\newcommand{\eea}{\end{eqnarray}}

\newcommand{\IS}{\bf S}

\newcommand{\ofa}{OF1_A} \newcommand{\ofb}{OF1_B}
\newcommand{\opa}{OP1_A} \newcommand{\opb}{OP1_B}

\preprint{CALT-68-2314\\ CITUSC/00-068 \\{\tt hep-th/0102095}}

\title{Strings on Orbifold Lines}

\author{Oren Bergman
\\
California Institute of Technology, Pasadena CA 91125, USA
\\ and\\
CIT/USC Center for Theoretical Physics \\
Univ. of Southern California, Los Angeles CA \\
\email{bergman@theory.caltech.edu} }
\author{Eric Gimon  and Barak Kol
\\
School of Natural Sciences \\ Institute for Advanced Study \\
Einstein Drive, Princeton NJ 08540,
USA\\
\email{gimon@sns.ias.edu}, \email{barak@sns.ias.edu} }

\abstract{The orbifold lines IIA/${\cal I}_8$ and IIB/${\cal I}_8
(-1)^{F_L}$ possess BPS discrete torsion variants which carry
fundamental string (NSNS) charge. We show that these variants are
actually classified by an integral electric field $F$ from the
twisted RR sector, and compute their tension and NSNS charge as a
function of $F$. The analysis employs equivariant K-theory and
the string creation phenomenon. The K-theory results demonstrate
the corrections to cohomology in the case of torsion;
 it is found that 8 units of $F$
are invisible at transverse infinity for IIA, and correspondingly
16 units for IIB.}

\begin{document}

\section{Introduction and Summary}
\label{intro}

Some string theory backgrounds are known to possess discrete
variants. In particular, certain orbifolds have discrete torsion
variants \cite{VafaTorsion,VW}, which can be thought of as arising
from finite discrete fluxes of the NSNS two form field $B$
concentrated at the fixed points. More recently it has been
realized that discrete RR fluxes can also give rise to orbifold
variants. In particular, the orbifolds $\R^8/{\cal I}_8$ and
$\R^8/{\cal I}_8(-1)^{F_L}$ of Type IIA and Type IIB string
theory, respectively, admit a set of variants corresponding to the
possible discrete RR fluxes at infinity. These fluxes were
originally classified in integer cohomology, and found to
correspond to torsion elements \cite{HananyKol}.

It has also been proposed recently that RR fields take values in
K-theory rather than integer cohomology \cite{MooreWitten}. It is
not surprising that RR charges, or D-branes, are described by
K-theory \cite{Moore_Minasian,WittenK}, since one needs to
specify a gauge bundle on the D-brane, as well as the homology
cycle it wraps, to fully characterize it. On the other hand
D-branes are sources for RR fields, so it is natural that the
latter should be valued in K-theory as well. Since K-theory
differs from cohomology only in
discrete torsion, it
becomes relevant for the study of the above types of variants.

This paper was partly motivated by the study of discrete torsion
variants of orientifolds \cite{HananyKol}(see also
\cite{Gimon,Hori,SethiO2}). In this approach, the choice of
orientifold projection leading to an $SO$ or $Sp$ gauge group
corresponds to a discrete torsion of the $B$ field. The objective
of \cite{HananyKol} was to list all possible variants by
identifying the background fields which allow for discrete
torsion. However, the analysis there relied on cohomology, and
some questions were left unsettled. In particular, questions were
raised about the variants of the above orbifold lines (which are
related to orientifolds by dualities), which were denoted $\ofa$
(IIA on $\R^8/{\cal I}_8$) and $\ofb$ (IIB on $\R^8/{\cal
I}_8(-1)^{F_L}$). These were found to have some variants which
carry fundamental string (NSNS) charge (hence the 'F' in their
names), but a general formula relating the RR fluxes to the NSNS
charge was not obtained. The main purpose of this paper is to
derive precisely such a relation.

We begin in section \ref{review} by reviewing the results of
\cite{HananyKol} on the discrete torsion variants of the above
orbifolds. For $\ofa$, the RR fluxes
were enumerated in reduced cohomology by
$H_R^{even}(\RP{7})=\Z_2\oplus\Z_2\oplus\Z_2$,\footnote{Reduced
cohomology is defined as
$H_R^*\equiv H^*/H^0=H^*/\Z$.}  and
were interpreted as arising from intersections with a fractional
D2-brane, D4-brane and D6-brane. The
intersection point serves as a ``domain wall'' on the orbifold
line, across which its tension, and accordingly its fundamental
string charge, can jump. The tension jump for the
three cases is given as follows
\be
 \begin{tabular}{|c|c|c|c|}
 \hline
 Brane & D6 & D4 & D2 \\
 \hline
 Tension jump & $1/16$ & $1/4$ & $1$ \\
 \hline
 \end{tabular} \;.
 \label{phenoma1}
\ee
We will review the derivation of these results, and explain how
they are consistent with T-duality. The review also includes some
basic examples of discrete torsion for completeness.

We then explain how re-interpreting RR fields in K-theory corrects
this picture. In particular, for the IIA orbifold, reduced
cohomology is replaced by the reduced K-theory group
$\widetilde{K}_{\Z_2}({\IS}^{8,0})=\widetilde{K}(\RP{7})=\Z_8$.
The ``enhancement'' of the discrete torsion group from cohomology
to K-theory actually has an interesting physical interpretation.
While in cohomology it appears that two fractional D$p$-branes
have a trivial discrete torsion, in K-theory we see that two
fractional D$p$-branes correspond to the same discrete torsion as
one D$(p-2)$-brane. So, for example, adding two D4's gives a D2
class instead of a trivial one. K-theory also allows us to make a
distinction between RR fields at infinity, which take values in
$\widetilde{K}_{\Z_2}({\IS}^{8,0})= \Z_8$, and RR fields in the
total (singular) transverse space, which take values in
$\widetilde{K}_{\Z_2}(\R^{8,0})= \Z$.\footnote{The latter space
is usually not analyzed by ordinary cohomology (``ordinary'' here
is ``non equivariant''), since it is singular.} The two groups
are related, and as a consequence the generator $x$ of the latter
group satisfies that $8x$ is invisible at
infinity.\footnote{Invisible at infinity means that it is mapped
to zero by the natural mapping into
$\widetilde{K}_{\Z_2}({\IS}^{8,0})$.} The analogous groups in the
IIB orbifold are $K^{-1}_{\pm}({\IS}^{8,0})=\Z_{16}$ and
$K^{-1}_{\pm}(\R^{8,0})=\Z$, so 16 times the generator of the
latter group is trivial at infinity.

In section \ref{perturbative} we analyze in detail the orbifolds
$\R^8/{\cal I}_8$ and $\R^8/{\cal I}_8(-1)^{F_L}$ for both Type
IIA and Type IIB. We determine both the perturbative closed
string spectrum and the D-brane spectrum of these theories. In
particular, it is shown that the ${\cal I}_8$ orbifold of IIA
($\ofa$) and the ${\cal I}_8(-1)^{F_L}$ orbifold of IIB ($\ofb$)
contain a massless vector field $A$ in the twisted RR sector.
Since this field lives in two dimensions it has no dynamics, but
it does allow for charged point-like defects, across which the
scalar field strength $F=*dA$ jumps by a discrete amount. These
defects are precisely the fractional D$p$-branes of $\ofa$ and
$\ofb$ (and a non-BPS D-particle in $\ofb$). The corresponding
jump in the value of $F$ is given by
 \be \label{Fcharge} \Delta
F_p = 2^{-p/2} \ee for the fractional D$p$-branes, and $\Delta
F_{\widetilde{0}}=\sqrt{2}$ for the non-BPS D-particle.

These jumps can now be identified with the integral charges in
$\widetilde{K}_{\Z_2}(\R^{8,0})$ (and $K^{-1}_{\pm}(\R^{8,0})$)
above, since both are generated by intersections with fractional
D-branes. In the IIA case, the generator $x$ of
$\widetilde{K}_{\Z_2}(\R^{8,0})$ corresponds to the fractional
D6-brane, which carries a charge $\Delta F=1/8$ , and the element
$8x$ is equivalent to the fractional D-particle, which has a
charge $1$, and is invisible at infinity. In the IIB case, the
generator corresponds to the fractional D7-brane (with charge
$\sqrt{2}/16$), and the element $16x$ is the non-BPS D-particle
(with charge $\sqrt{2}$). This is an example of how K-theory
correlates different RR fields; in this case the different fields
are the twisted sector and bulk RR fields.

Section \ref{Tension and Charge} contains our main result, namely
the tension and NSNS charge of the orbifold lines as a function of
the twisted sector scalar field $F$. We find that \be
\label{phenomQ}
 Q (F)=4 F^2  - {1 \over 16} \;,
\ee which together with (\ref{Fcharge}) is consistent with the
earlier results (\ref{phenoma1}). The constant $Q(0)=-1/16$ was
computed in \cite{SenOrbifolds} based on the term $\int B \wedge
Y_8(R)$ in the effective Type IIA action \cite{VW1loop} (see
section \ref{perturbative} for more detail, and \cite{SVW} for
related work). This value ties well with the known charges of
objects related by dualities to orbifold lines; both the
orientifold line and the $OM2^-$ have the same
tension\footnote{$OM2^-$ denotes $M/{\cal I}_8$.} (for the $OM2$
see \cite{SethiO2}). Our concern here is to derive the $F$
dependence of $Q(F)$, that is to derive $\Delta Q(F)=Q(F)-Q(0)$.

The relation (\ref{phenomQ}) is first obtained from the low-energy
effective action. The kinetic term for $F$ gives a vacuum energy
per unit length, {\em i.e.} a tension, proportional to $F^2$. The
correct normalization is determined by the sphere closed string
diagram with two insertions of $F$, which is related to a simple
one-loop open string amplitude that computes the supersymmetry
index of the internal ($\R^8/\Z_2$) CFT.

In an alternative approach, we use the fact that pairs of
fractional D-branes in these orbifolds are linked, and therefore
exhibit the phenomenon of string creation
\cite{HananyWitten,StringCreation,BGL}. We consider specifically
the case of two fractional D-particles in $\ofa$. This system is
similar to the D0-D8 system in ten dimensions, in which a single
string is created when the branes cross. The number of strings
created can be understood in terms of the supersymmetric index of
the 0-8 open string. Our case is a bit more subtle, in that the
particle and domain wall are both D-particles, but the number of
strings which are created is still given by the supersymmetry
index, which in this case is $n=8$. The number of strings ending
on a single D-particle, and therefore the tension jump of the
orbifold line, can then be determined from a consistency condition
for the process of exchange, and the result is consistent with
(\ref{phenomQ}).

Note that the constant $n=8$ is determined in two seemingly
unrelated ways - the first from K-theory, and the second as an
open string index.
We argue in the same section that the two are actually related, by
considering a source term for $B_{NS}$ in the Type IIA action,
which depends on a certain self-intersection number. Here we
assume that $\R^8/ \Z_2$ has a singular (shrunk) 4-cycle $C$, such
that the fractional D-particle is a D4-brane wrapping $C$, and
that its self-intersection number must be $\pm 8$. On one hand the
self intersection number is geometric and determined by K-theory,
but on the other hand when two such D4-branes are exchanged the
number of strings created is exactly the self-intersection number,
so the two approaches are connected.

The case of $\ofb$ is similar to $\ofa$, and the reader may
concentrate his / her attention on the latter in a first reading.
Collecting the results for the $\ofb$ we find that the K group at
infinity is $\Z_{16}$, replacing the cohomological $(\Z_2)^4$
which arises from intersections with the fractional D7, D5, D3,
and D1. The normalization for $F$ is that the fractional D7 (the
generator) has charge $\sqrt{2}/16$, while the non-BPS D-particle,
which is invisible at infinity, has charge $\sqrt{2}$. The
formulas (\ref{Fcharge}) and (\ref{phenomQ}) are the same for both
orbifold lines.

\section{RR discrete torsion}
\label{review}

\subsection{Discrete torsion}
\label{Torsion}

It was observed in \cite{VafaTorsion} that for certain orbifolds
one can multiply the twisted components of the string partition
function by phases, while maintaining consistency, and in
particular modular invariance. For an orbifold $M/\Gamma$, where
is $M$ is the covering space on which a discrete group $\Gamma$
acts, the possibilities for adding such phases were found to be
classified by the second cohomology group of $\Gamma$ with
$U(1)$ coefficients $H^2(\Gamma,U(1))$. The simplest example is
$H^2(\Z_n \times \Z_n,U(1))=\Z_n$.

Alternatively, discrete torsion can be viewed as arising from {\it
spacetime} cohomology rather than {\it group} cohomology (see
\cite{Sharpe} for a rigorous treatment). The phases are due to a
non-trivial topology of the NSNS $B$ field. The contribution of
$B$ to the functional integral is through the term $\exp({i \over
2}\int B_{\mu\nu}
\partial_{\alpha} X^\mu
\partial_{\beta} X^\nu \epsilon^{\alpha \beta})=\exp({i \over 2}
\int B^{induced})$, and therefore in the presence of a flat $B$
field all world-sheets in the same spacetime homology class will
receive the same contribution. Thus, by definition, the phases are
classified by the cohomology $H^2(M/\Gamma,U(1))$. This can be
further simplified by observing that the sequence $0 \to \Z \to
\R \to U(1) \to 0$  identifies the components of
$H^2(M/\Gamma,U(1))$ with the discrete torsion part of
$H^3(M/\Gamma,\Z$). It is natural to identify this class with the
class of the field strength $[H]=[dB] \in H^3(M/\Gamma,\Z)$.

By analogy, if we replace the world-sheet of the fundamental string
with the world-volume of a D-brane, we expect that the full
non-perturbative partition function will have additional discrete
torsion degrees of freedom. Namely, for any RR $p$-form field
strength $G^{(p)}$, one expects to see variants classified by
$H^p(M/\Gamma,\Z)$. For Type IIA this implies
$H^{even}(M/\Gamma,\Z)$ variants, while for IIB
$H^{odd}(M/\Gamma,\Z)$ variants are expected. We will soon see
that this picture is in fact corrected when we re-interpret RR
fields, and in particular RR discrete torsion, in K-theory.

The simplest example of discrete torsion corresponds to a vector field
on $\R^d / \Z_2$, where the $\Z_2$ acts as $x \to -x$.
In this case $H^2((\R^d \backslash \{ 0 \})/
\Z_2,\Z)=\Z_2$, so a non-trivial configuration for the gauge
field is possible. Such a configuration would be detected by
sending a charged particle on a closed path from a point to its
antipode, and observing a phase of $-1$, even though the field
strength is locally zero. In other words we have a flat circle
bundle $(\R^d \times {\IS}^1)/ \Z_2$, with the $\Z_2$ acting on the
circle $0 \le \theta \le 1$ by $\theta \to \theta + 1/2$.

\subsection{Previous results from cohomology}

Let us review the results obtained in \cite{HananyKol} regarding
the classification and properties (namely the tension and NSNS
charge) of the orbifolds $\ofa$ and $\ofb$. The relevant
cohomologies for $\ofa$ are $H^{even}(\RP{7},\Z)=H^0 \oplus H^2
\oplus H^4 \oplus H^6= \Z \oplus \Z_2 \oplus \Z_2 \oplus \Z_2$,
corresponding respectively to the RR fields $G^{(0)}, G^{(2)},
G^{(4)}$, and $G^{(6)}$. In $\ofb$ all the RR forms are twisted
in the sense that they switch sign upon inversion. The relevant
cohomologies are therefore
$H^{odd}(\RP{7},\widetilde{\Z})=\widetilde{H}^1 \oplus
\widetilde{H}^3 \oplus \widetilde{H}^5 \oplus \widetilde{H}^7 =
\Z_2 \oplus \Z_2 \oplus \Z_2 \oplus \Z_2$, corresponding to
$G^{(1)},\ldots,G^{(7)}$, where $\widetilde{\Z}$ is the twisted
bundle (or sheaf) of integers on $\RP{7}$.

Discrete torsion fluxes can be interpreted as arising from
fractional branes. In order to create a $\Z_2$ flux for a $p$-form
field strength $G^{(p)}$ one takes a D$(8-p)$-brane which is its
source, and moves it onto the orbifold where the D-brane can
split into two halves. By sending the two halves far away in
opposite directions one constructs a new orbifold which carries
the required discrete torsion.

According to this interpretation $\ofa$ can be intersected by D8,
D6, D4 and a D2. The D8 is special as it changes the
cosmological constant. It is represented by the $\Z$ factor in
$H^{even}(\RP{7},\Z)$. The $\ofb$ on the other hand can be
intersected by the D7, D5, D3 and D1. Note that all of the above
intersection configurations are supersymmetric, and that in all
cases the flux from a jump in the fundamental string charge can be
carried away by fields in the world-volume of the brane.

Let us see how the jump in fundamental string charge was
determined in some cases by an ad-hoc series of dualities.
\begin{itemize}
\item The intersection of the $\ofa$ with a D4-brane
lifts in M theory to an
intersection in 11d of the OM2 with an M5-brane. This is known
to cause a jump of $+1/4$ in the membrane charge \cite{SethiO2},
which upon compactification becomes the fundamental string charge.
\item The intersection of the $\ofb$ with a D5-brane is S-dual to an
intersection of the orientifold line with an NS5-brane. This
intersection is known to change an $O1^-$ ($SO$ projection) into an
$O1^+$ ($Sp$ projection),
causing a jump of $1/16-(-1/16)=+1/8$ in the D-string charge,
which is mapped into the F-string charge in the original picture.
\item  The intersection with the D3 can be analyzed in the same way:
after S-duality one gets an orientifold line $O1^-$ which is
transformed by the D3-brane into an $\widetilde{O1}^-$,
which is an $O1^-$ with an extra half D1-brane. Therefore the jump
here is by $+1/2$.
\item  In agreement with the trend, one also expects the intersection with
a D6 to lead to a tension jump of $+1/16$. For that one needs to
assume that the tension of the ``bare'' $\ofa$ is $-1/16$ (see
section \ref{perturbative}). The intersection with a D6 is
equivalent to a discrete torsion for the vector field of Type IIA.
When lifted to M theory, it is described by a smooth manifold
$(\R^8 \times {\IS}^1)/ \Z_2$ (as explained in the example in
previous subsection), and is not expected to carry any charges.
\item For the D2-brane it was found by a less direct method that
the jump is $+1$.
\end{itemize}
We summarize the results in the following table
\be
 \begin{tabular}{|c|c|c|c|c|c|}
 \hline
 Brane & D6 & D5 &D4 & D3 &D2 \\
 \hline
 Tension jump, $\Delta Q$ & $1/16$ & $1/8$ & $1/4$ & $1/2$ & $1$ \\
 \hline
 \end{tabular} \;.
\ee The pattern that emerges can be explained by T-duality.
Consider the orbifold to act on a torus, that is ${\bf T}^8 / \Z_2$,
rather than $\R^8 / \Z_2$ as we did so far. In this case the
number of fixed points on the torus is $2^8=256$. When we perform
a T-duality in one of the directions of the torus, the 256 $\ofa$
lines are replaced by 256 $\ofb$ lines (or vice versa), and
D-branes are replaced according to the usual rule. Since the
total tension jump (the sum of jumps for all 256 lines) is
invariant under T-duality, and since a D$p$-brane intersects
$2^p$ lines, it follows that the jump on a single orbifold line is
proportional to $2^{-p}$, in agreement with the table.

\subsection{K-theory corrections}
\label{K-theory}

It has become clear in the past year that RR fields and charges
take values in K-theory, rather than integer cohomology. This was
suggested by the fact that the states which carry RR charges,
namely D-branes, possess an internal structure which includes a
gauge bundle \cite{Moore_Minasian},
and from the conjecture that a brane and an anti-brane which carry
isomorphic bundles annihilate into the vacuum
\cite{Sen_conjecture,WittenK}. It follows that D-branes are
classified by isomorphism class groups of gauge bundles, {\em
i.e.} K-theory. In particular, D-branes on a smooth space $X$ are
classified by the compactly supported K-theory groups
$K^{-1}_{cpct}(X)$ and $K_{cpct}(X)$ in Type IIA and Type IIB
string theory, respectively.\footnote{To be precise, consistent
(tadpole-free) RR charges in Type IIB are actually given by the
kernel of the natural map $K_{cpct}(X) \rightarrow K(X)$ (with the
obvious generalization to Type IIA).} Moreover, since D-branes are
sources of RR fields, the latter also take values in K-theory
\cite{MooreWitten}. In particular, source free RR fluxes in a
space $Y$ are classified by $K(Y)$ in Type IIA, and by $K^{-1}(Y)$
in Type IIB.

K-theory is also well-suited to handle singular spaces such as
orbifolds and orientifolds, and it is in these cases where
K-theory generally differs from cohomology. In orbifold
backgrounds RR charges and fields take values in {\em equivariant}
K-theory. In particular , for orbifolds of the type $\R^n/{\cal
I}_n$ the relevant groups are $K_{\Z_2}$ and $K^{-1}_{\Z_2}$, and
for orbifolds of the type $\R^n/{\cal I}_n(-1)^{F_L}$ the
corresponding groups are $K_{\pm}$ and $K^{-1}_{\pm}$.

For the orbifolds $\ofa$ and $\ofb$
K-theory differs from cohomology in two ways. First, RR fluxes at
infinity take values in\footnote{The notation ${\IS}^{l,m}$ refers
to the sphere at infinity in the space $\R^{l,m}$, which corresponds
to $\R^{l+m}/\Z_2$, where the $\Z_2$ inverts $l$ coordinates.
In particular ${\IS}^{l,0}\sim \RP{l-1}$, but the former notation
reminds us that the $\Z_2$ acts on the bundles as well.}
 \be \label{infinityRR}
\begin{array}{rcll}
 \widetilde{K}_{\Z_2}({\IS}^{8,0}) & = &
     \Z_8 & \;\; \mbox{for}\;\; OF1_A \\[5pt]
 K^{-1}_{\pm}({\IS}^{8,0}) & = &
     \Z_{16} & \;\; \mbox{for}\;\; OF1_B \;.
\end{array}
\ee
The discrete torsion corresponding to the fields $G^{(8-p)}$,
with $1\leq p\leq 7$, is different from cohomology.
Recall that in cohomology the
torsion subgroup was $(\Z_2)^3$ and $(\Z_2)^4$ for $OF1_A$ and
$OF1_B$, respectively. The dimensions agree with K-theory, but the
group structure is different. Let us denote the $\Z_2$ generator
in the $n$'th cohomology group by $x^{(n)}$, where $n=2,4,6$ for
$OF1_A$, and $n=1,3,5,7$ for $OF1_B$. In cohomology these satisfy
\be
2x^{(n)}=0\;.
\ee
  This can be understood physically by the fact that two fractional
D$p$-branes (with $p\leq 7$) can combine into a bulk D$p$-brane,
which can separate from the orbifold line, and thereby remove the
flux from the $\RP{7}$ at infinity. On the other hand, the above
results suggest loosely that in K-theory \be \label{Ktorsion}
2x^{(n)}=x^{(n+2)} \ee (see the Appendix for a more rigorous
derivation). Physically this means that the RR flux at infinity
due to two fractional D$p$-branes ($p\leq 7$) is not trivial, but
rather corresponds to the RR flux of a fractional
D$(p-2)$-brane.\footnote{Two comments are in order here. First,
for $\ofa$ there is also an integral $G^{(0)}$ flux, which is
apparent in the unreduced group $K_{\Z_2}({\IS}^{8,0}) = \Z \oplus
\Z_8$, and which agrees with unreduced cohomology. However, the
cohomology generator satisfies $2x^{(0)}=\{G^{(0)}=2\}$, whereas
the K-theory generator satisfies $2x^{(0)}=\{G^{(0)}=2\} +
x^{(2)}$. The second comment is that torsion group structure in
K-theory implies that each fractional D$p$-brane carries $1/2$
unit of fractional D$(p-2)$-brane charge, which in turn requires
an anomalous Dirac quantization condition for the gauge field on
the $p$-brane
$$
 \int_{\RP{2}\subset\RP{7}} {F\over 2\pi} =
{1\over 2} + \Z\;.
$$
The latter can be viewed as a manifestation
of the difference between K-theory torsion and cohomology
torsion.} We could still recover a bulk D-brane, however, by
combining two fractional D-branes with opposite twisted charge
(corresponding to $x^{(n)}$ and $- x^{(n)}$) if they carry the
same bulk charge.

The second K-theory correction appears in the spectrum of RR
fields in the total transverse space, specifically when we compare
fields which are compactly supported to the more general case. The
former take values in \be \label{compactRR}
\begin{array}{rcll}
 K_{\Z_2,cpct}(\R^{8,0}) & = &
     \Z \oplus \Z  & \;\; \mbox{for}\;\; OF1_A \\[5pt]
 K^{-1}_{\pm,cpct}(\R^{8,0}) & = &
     \Z & \;\; \mbox{for}\;\; OF1_B \;.
\end{array}
\ee
Since the only compactly supported RR fields are associated with
D-particles, it is clear that one of the integral fluxes in $\ofa$
corresponds to $G^{(8)}$. As we shall see in the next section, the
other integral flux, as well as the flux in $\ofb$, corresponds to
a twisted sector RR field $F$, under which all fractional D-branes
(as well as a stable non-BPS D-particle in $\ofb$) are charged.
Both of these fluxes are
absent on the $\RP{7}$ at infinity (\ref{infinityRR}); the 8-form
cannot be supported on a 7-manifold, and the twisted sector field
is supported only at the origin. On the other hand, the fluxes which
were visible at infinity are absent here because they
are not compact in the transverse space. If we relax the condition
of compact support we get
\be
\label{generalRR}
\begin{array}{rcll}
 \widetilde{K}_{\Z_2}(\R^{8,0}) & = & \Z & \;\; \mbox{for}\;\;
  OF1_A \\[5pt]
 K^{-1}_{\pm}(\R^{8,0}) & = & \Z & \;\; \mbox{for}\;\; OF1_B \;.
\end{array}
\ee
This is surprising at first, since one expects all the fluxes
to contribute in this case, but clearly (\ref{generalRR}) is not
the direct sum of (\ref{infinityRR}) and (\ref{compactRR}).
In fact, the three groups are part of a long exact sequence (Appendix),
which shortens to
\be
\begin{array}{ccccccccccc}
 0 & \rightarrow & K_{\Z_2}^{-1}({\IS}^{8,0}) & \rightarrow &

 K_{\Z_2,cpct}(\R^{8,0}) &
 \stackrel{i}{\rightarrow} &
 \widetilde{K}_{\Z_2}(\R^{8,0}) & \rightarrow &
 \widetilde{K}_{\Z_2}({\IS}^{8,0}) & \rightarrow & 0 \\
  & & \parallel & & \parallel & & \parallel & & \parallel & & \\
   & & \Z' & & \Z'\oplus\Z_y &
  \stackrel{\times 8}{\longrightarrow} &
  \Z_x & & \Z_8 & &
\end{array}
\ee
for $OF1_A$, and to
\be
\begin{array}{ccccccccc}
 0 &  \rightarrow &
 K_{\pm,cpct}^{-1}(\R^{8,0}) &
  \stackrel{i}{\rightarrow} &
 K_{\pm}^{-1}(\R^{8,0}) & \rightarrow &
 K_{\pm}^{-1}({\IS}^{8,0}) & \rightarrow & 0 \\
   & & \parallel & & \parallel & & \parallel & & \\
   & & \Z_y & \stackrel{\times 16}{\longrightarrow} & \Z_x & & \Z_{16} & &
\end{array}
\ee for $OF1_B$.\footnote{In the first sequence the first two K
groups are actually equal to their reduced versions.} We have
labeled the different integral fluxes according to how they map in
the sequence. It follows that $\Z$ and $\Z'$ are identified with
$F$ and $G^{(8)}$, respectively.\footnote{Note that the flux of
$G^{(8)}$ ($\Z'$) is present in $K_{\Z_2,cpct}(\R^{8,0})$, but
absent in $\widetilde{K}_{\Z_2}(\R^{8,0})$. This should follow
from an argument analogous to the one given in \cite{MooreWitten}
for the triviality of D-brane charge in non-compactly supported
K-theory (see footnote~6).} The sequences show that the generator
$y$ of $\Z$ in the compactly supported group is related to the
generator $x$ of $\Z$ in the general group by \be
\begin{array}{rcll}
\label{Kproduct}
 y &=& 8x & \;\; \mbox{for}\;\; OF1_A \\[5pt]
 y &=& 16x & \;\; \mbox{for}\;\; OF1_B \;.
\end{array}
\ee The element $y$ corresponds to the $F$-flux of a fractional
(stable non-BPS) D-particle in $OF1_A$ ($OF1_B$), and is
therefore trivial on the $\RP{7}$ away from the orbifold line. On
the other hand $x$ maps to the generator of the torsion group on
$\RP{7}$, and therefore corresponds to the $F$-flux of a
fractional D6-brane (D7-brane) in $OF1_A$ ($OF1_B$). It then
follows that the element $2^k\cdot x$ (for $k \le 3$) of the
non-compactly supported group corresponds to the $F$-flux of the
fractional D$(6-2k)$-brane (D$(7-2k)$-brane). The relative
$F$-fluxes of the fractional D$p$-branes are therefore given by
\be
 F \propto 2^{-p/2}\;.
\ee
We will make this more precise in the following section.

\section{Perturbative orbifold lines}

\label{perturbative}

In this section we analyze in detail the
perturbative $\Z_2$ orbifold lines $\R^{1,1}\times \R^8/{\cal
I}_8$ and $\R^{1,1}\times \R^8/{\cal I}_8(-1)^{F_L}$ for both Type
IIA and Type IIB string theory. The Type IIA orbifolds are denoted
$\ofa$ and $\opa$, respectively, and the Type IIB orbifolds are
denoted $\opb$ and $\ofb$, respectively. The four theories are related
by T-duality if we compactify some of the directions. In
particular, $OF1_A$ is related to $OF1_B$ by T-duality of a
coordinate along $\R^8$, and to $OP1_B$ by T-duality
along the line. These orbifolds preserve 1/2 of the supersymmetry
of Type II string theory. In particular, from the point of view of
the invariant subspace $\R^{1,1}$, $OF1_{A,B}$ possess
${\cal N}=(8,8)$ supersymmetry, and $OP1_{A,B}$ have ${\cal
N}=(16,0)$ (or $(0,16)$) supersymmetry.

\subsection{Perturbative spectrum}

We begin by describing the closed string spectrum of these
theories. The {\em untwisted sector} is obtained simply by acting
on the spectrum of Type II string theory by the orbifold
projection
\begin{equation}
\label{projection}
 P={1\over 2}(1+g)\;, \quad (g={\cal I}_8\;\;
  \mbox{or}\;\;{\cal I}_8(-1)^{F_L}) \;.
\end{equation}
Before orbifolding, the Type II spectrum in light-cone gauge
consists of states created from a ground state in one of four
sectors (NS-NS, NS-R, R-NS, and R-R) by left and right-moving
oscillators of the form $\alpha^i_{-n}$, $\psi^i_{-n}$ (R),
$\psi^i_{-n+1/2}$ (NS), where $n\geq 0$, and $i=1,\ldots, 8$ (the
fixed line is taken to lie along the longitudinal coordinate
$x^9$). The ground state of the NS sector is non-degenerate and
tachyonic, with $m^2=-1/2$, whereas the presence of 8 fermionic
zero modes in the R sector give rise to a 16-fold degenerate
massless ground state, {\em i.e.} a massless spinor. Consistency
(and spacetime supersymmetry) requires that this spectrum be
projected by $P_{GSO}\cdot \widetilde{P}_{GSO}$, where
\begin{equation}
 P_{GSO}=\left\{
 \begin{array}{rl}
  {1\over 2}(1+(-1)^f) & \mbox{NS}\\[3pt]
  {1\over 2}(1+(-1)^f) & \mbox{R}
 \end{array}\right.
 \qquad
 \widetilde{P}_{GSO}=\left\{
 \begin{array}{rl}
  {1\over 2}(1+(-1)^{\widetilde{f}}) & \mbox{NS}\\[3pt]
  {1\over 2}(1\pm (-1)^{\widetilde{f}}) & \mbox{R} \;.
 \end{array}\right.
\end{equation}
In the last line the $+$ sign corresponds to Type IIB and the $-$
sign to Type IIA. In particular, the projection on the R ground
state reduces it to a chiral 8-component spinor (either ${\bf
8}_s$ or ${\bf 8}_c$ of $SO(8)$).

In the {\em twisted sector} the moding of the oscillators in the
inverted directions changes, and becomes half-odd-integral for the
bosons $\alpha^i_{-n+1/2}$ and Ramond fermions $\psi^i_{-n+1/2}$
(R), and integral for the Neveu-Schwarz fermions $\psi^i_{-n}$
(NS). The ground state of the R sector, as usual, is massless, due
to a cancellation between the bosonic and fermionic contributions.
Now this state is a singlet of $SO(8)$,
since there are no fermionic zero modes. The ground state of the
NS sector on the other hand is 16-fold degenerate, but is massive,
with $m^2=+1/2$. The only massless field is therefore a real RR
scalar $\phi$. Since the $SO(1,1)$ chirality is fixed in
light-cone gauge, this is actually a {\em chiral boson}.

A-priori this field exists in all four cases, but supersymmetry
only allows it in the ${\cal N}=(16,0)$ theories. We therefore
expect it to be removed from the other two theories by the GSO
projection. This is most easily seen in a covariant gauge, where
the $SO(1,1)$ chirality is not fixed. The R ground state is then a
two-component Majorana spinor of $SO(1,1)$, which decomposes as
${\bf 2}={\bf 1}_{+1/2} \oplus {\bf 1}_{-1/2}$. The spectrum of
fields in the twisted RR sector is then given by
\begin{equation}
 {\bf 2}\times {\bf 2} = {\bf 1}_{+1} \oplus {\bf 1}_{-1}
        \oplus 2 \cdot {\bf 1}_0\;.
\end{equation}
The first term corresponds to a chiral (self-dual) boson $\phi^+$,
and the second to an anti-chiral (anti-self-dual) boson $\phi^-$.
The third term is an unphysical state, and is removed in
light-cone gauge. The GSO projection in the twisted sector of the
${\cal I}_8$ orbifold is the same as in the untwisted sector above, which
means that the chiral and anti-chiral bosons are removed from Type
IIA, but one of them remains in Type IIB. In the twisted sector of
the ${\cal I}_8(-1)^{F_L}$ orbifold the left-moving GSO projection is reversed
relative to the untwisted sector, so one of the bosons remains in
Type IIA, and both are removed from Type IIB. We therefore find a
massless chiral (or anti-chiral) RR boson only in the twisted
sector of $OP1_{A,B}$. Finally, we must again project by
(\ref{projection}), but this does not affect the above state.

The orbifolds $OF1_{A,B}$ do not have any massless propagating
fields in the twisted sector. They may however possess
non-dynamical fields, which are consistent with ${\cal N}=(8,8)$
supersymmetry. An example in ten dimensions is the RR 9-form
potential in Type IIA string theory. This field is not dynamical,
but is crucial for describing massive Type IIA backgrounds in
which D8-branes are present. The analogous field in two dimensions
is a vector potential $A$. We claim that precisely such a field
exists in the twisted RR sector of the orbifolds $OF1_{A,B}$. In
support of this claim we shall offer in the next subsection
precisely the same argument which appeared in support of the
9-form in ten dimensions \cite{Polchinski}, namely the existence
of a corresponding D-brane, which in our case is a (fractional)
D-particle.

In the compact case, ${\bf T}^8/\Z_2$, there are additional fields
corresponding to 16 fundamental strings,
which must be added to cancel a one loop tadpole
of the form $\int B \wedge Y_8(R)$ \cite{VW1loop,SenOrbifolds}
(see also \cite{SVW}). This sets the NSNS charge (and tension)
of a single bare orbifold line to $-1/16$.
That the $\ofb$ has the same $Q(0)=-1/16$ is a consequence of T
duality (in some direction in the ${\bf T}^8$) which preserves
both the number of orbifold lines and the number of fundamental
strings needed to cancel the tadpole.
This charge is also
consistent with dualities: the $\ofa$ lifts in M theory to the
orbifold 2-plane (OM2), whose tension was determined to be $-1/16$
in \cite{SethiO2}; the $\ofb$ is S dual to the bare orientifold
line which has tension $-1/16$ as well. Later we will find out
that for the $\ofa$ one gets an orbifold with zero tension by
adding a D6-brane ($G^{(2)}$ torsion) to the bare (i.e.
perturbative) orbifold.  This lifts to a smooth orbifold in
M-theory of the form $(\R^8 \times {\IS}^1)/\Z_2$, where the $\Z_2$
acts as a half-shift along the M-circle.

\subsection{D-branes}

The analysis of the D-brane spectrum follows closely the boundary
state techniques of \cite{BG_D-particle,Gaberdiel_Stefanski}. We
will use the convention of defining a D$p$-brane of type $(r,s)$,
where $r+s=p$, to have $r$ spatial world-volume directions along
the orbifold line\footnote{$r=-1$ corresponds
to a Dirichlet boundary condition in time, {\em i.e.} a
D-instanton.}, and $s$ along $\R^8$. The
spectrum of D-branes on $\Z_2$ orbifolds contains two kinds of
irreducible D-brane states: {\em fractional} D-branes and {\em
truncated} D-branes. The former are linear combinations of
boundary states in the untwisted NSNS and RR sectors, and the
twisted NSNS and RR sectors,
\begin{equation}
\label{fractional}
 \ket{D(r,s)} = {1\over 2}\Big(
 \ket{B(r,s)}_{NSNS} + \epsilon_1\ket{B(r,s)}_{RR}
 + \epsilon_1\epsilon_2\ket{B(r,s)}_{NSNS,T}
 + \epsilon_2\ket{B(r,s)}_{RR,T}\Big)\;,
\end{equation}
and the latter contain only an untwisted NSNS state and a twisted
RR state,
\begin{equation}
\label{truncated}
 \widetilde{\ket{D(r,s)}} = {1\over\sqrt{2}}\Big(\ket{B(r,s)}_{NSNS} +
\epsilon_2\ket{B(r,s)}_{RR,T}\Big)\;.
\end{equation}
The numerical factors in front are fixed by the requirement that,
for any pair of D-branes, the cylinder amplitude corresponds to an
open string partition function \cite{Cardy}. The signs
$\epsilon_1,\epsilon_2$ determine the charges of the D-brane with
respect to the appropriate massless RR field in the untwisted and
twisted sector, respectively. The presence of twisted sector
charge means that these D-branes are ``stuck'' to the orbifold
line. By combining two fractional D-branes with opposite
$\epsilon_2$ one can form a {\em bulk} D-brane, which can separate
from the orbifold line.

In a supersymmetric orbifold, such as the ones we are considering,
fractional D-branes are BPS states, and truncated D-branes
describe non-BPS states. The latter may or may not be stable,
depending on whether their spectrum includes tachyons. In the
non-compact case we are considering, truncated D-branes will only
be stable for $s=0$, {\em i.e.} if their world-volume is entirely
transverse to the $\R^8$.

The spectrum of physical D-brane states is determined by requiring
GSO and orbifold invariance. This has been done for the general
case of $n$ inverted directions in \cite{Gaberdiel_Stefanski}. The
results for $n=8$ are summarized in the following table \be
\begin{tabular}{|l|l|l|l|l|}
\hline
 D-brane & $OF1_A$ & $OP1_A$ &
 $OP1_B$ & $OF1_B$ \\
\hline
 $\ket{D(r,s)}$ & $r=0$, $s$ even & $r=\pm 1$, $s$ odd
  & $r=\pm 1$, $s$ even & $r=0$, $s$ odd \\[5pt]
 $\widetilde{\ket{D(r,s)}}$ & -- & $r=\pm 1$, $s=0$ & -- & $r=s=0$\\
\hline
\end{tabular}
\label{Dbranes}
\ee
In particular both $OF1_A$ and $OF1_B$ admit a
D-particle with a twisted RR component. In the first case it is a
BPS fractional D-particle, and in the second case it is a stable
non-BPS truncated D-particle. This confirms the existence of a
non-dynamical vector field $A$ in the twisted sector of these
theories.

In addition, $OF1_A$ contains fractional D$p$-branes
with $p=2,4,6,8$, which are transverse to the orbifold line and
intersect it at a point. All of these are charged under $A$, as
well as under the respective RR field $C^{(p+1)}$ in the untwisted sector.
Similarly, $OF1_B$ contains transverse fractional
D$p$-branes with $p=1,3,5,7$, which are all charged under $A$, as
well as the respective $C^{(p+1)}$. The two-dimensional scalar
field strength $F=*dA$ can be thought of as a cosmological
constant in the twisted sector, by analogy with the
ten-dimensional cosmological constant $G^{(0)}=*dC^{(9)}$ in Type
IIA string theory. The above fractional branes form domain walls
on the $OF1$ orbifolds, across which the value of $F$ jumps. The
magnitude of the jump depends on the size of the fractional
D-brane as follows:
\begin{equation}
\label{RRjump}
 \Delta F_{Dp} = 2^{-p/2} \;.
\end{equation}
This is most easily determined by replacing $\R^8$ with the
compact ${\bf T}^8$. In this case there are $2^8$ fixed lines,
each containing a RR 1-form in the twisted sector. The correctly
normalized fractional D-brane states are (ignoring the volume
factors) \cite{Gaberdiel_Stefanski}
\begin{eqnarray}
\label{compact_fractional}
 \ket{D(r,s)} &=& {1\over 2} \bigg[
 \ket{B(r,s)}_{NSNS} + \epsilon_1\ket{B(r,s)}_{RR} \nonumber\\
  && \mbox{}+ 2^{-s/2}\sum_{i=1}^{2^s} \epsilon_2^{(i)}\Big(
     \epsilon_1\ket{B(r,s)}_{NSNS,T_i}
 + \ket{B(r,s)}_{RR,T_i} \Big)\bigg]\;,
\end{eqnarray}
where $i$ labels the orbifold lines which the D-brane intersects.
Equation (\ref{RRjump}) follows by noting that $s=p$ for the $OF1$
fractional D-branes. In particular, for the fractional D-particle
in Type IIA the jump is $\Delta F =1$. On the other hand, for the
truncated D-particle in Type IIB the jump is
\begin{equation}
 \Delta F_{\widetilde{D0}} = \sqrt{2} \;.
\end{equation}
This follows by comparing its boundary state (\ref{truncated}) to
that of the Type IIA fractional D-particle (\ref{fractional}).
Alternatively, this also follows from the realization of the
non-BPS D-particle as a bound state of two BPS D-strings with
opposite untwisted RR charge and equal twisted RR charge
\cite{Sen_nonBPS}.

\section{Orbifold line tension and NSNS Charge}
\label{Tension and Charge}

The perturbative orbifold lines (with $F=0$) carry both an NSNS
charge and tension given by $Q(0)=-1/16$. That the two are equal
is guaranteed by the fact that the orbifold lines are BPS
objects, {\em i.e.} they preserve 1/2 of the underlying
supersymmetry.\footnote{More precisely they preserve 16
supersymmetries.}
 This will continue to be true for non-zero
values of $F$, since $F$ does not have a fermionic partner, and
therefore does not appear in the SUSY variation equations. As we
shall see, the tension, and therefore the charge, depends on the
value of $F$.

\subsection{Effective action}

The low-energy effective action of the orbifold theories contains
a standard kinetic term for the twisted sector vector field
\be
\label{pert tension}
 S_F = - {n\over 2} \int F^2  \;,
\ee where the coefficient $n$ remains to be determined, though a
$1/2$ prefactor is inserted with hindsight. This implies that the
tension jump due to a D$p$-brane is given by $\Delta Q_{Dp}=(n/2)
\Delta F_{Dp}^2$. The normalization factor $n$ could a-priori be
different for IIA and IIB, but T-duality actually shows it is the
same. Consider a fractional D$p$-brane in IIA intersecting ${\bf
T}^8/ \Z_2$. Since a fractional D$p$-brane intersects $2^p$
orbifold lines, the total tension jump is $\Delta Q=2^p (n_A/2)
\Delta F_{Dp}^2=n_A/2$. After T-duality the same computation yields
$\Delta Q=n_B/2$, but since tension is invariant under T-duality
we have $n_A=n_B$.

To determine $n$ we could in principle compute
the closed string sphere amplitude with two insertions of $F$.
Alternatively, it can also be deduced from the cylinder amplitude
involving the twisted RR component of a fractional D-particle,
\be
 A_{RR,T} = {1\over 4}\bra{B0}
 e^{-lH_{closed}}\ket{B0}_{RR,T} \;,
\ee
which in the limit $l\rightarrow\infty$ is equivalent to the
sphere amplitude. On the other hand, the cylinder amplitude can
be re-expressed as a one loop open string amplitude, which in the
above limit reduces to the supersymmetry index of the internal
CFT, giving
\be \label{index}
 n = - \mbox{Tr}^{0-0}_{NS-R}(-1)^f\left({1+g\over 2}\right) = 8 \;.
\ee

The computation proceeds as follows. There are 8 massless states
in the NS sector $\psi_{-1/2}^{i}\ket{0}_{NS}$ prior to the
projection, where the index $i=1,..,8$ is transverse to the $\ofa$.
However since $\psi_{-1/2}^{i} $ is odd under the orbifold
projection, all these states are projected out, and no spacetime
bosons remain. In the R sector there are 8 fermionic zero modes
$\psi_0^{i}$, which upon quantization yield 16 massless states.
The action of $g$ on the R ground state is given by $\prod_i
\psi_0^{i}$, {\em i.e.} it is the same as the action of $(-1)^f$
on the R ground state in the internal CFT, so the only
non-vanishing contribution comes from
${1\over 2}\mbox{Tr}_R(-1)^fg={1\over 2}\mbox{Tr}_R{\bf 1}=8$.

The tension jump is therefore given by
\be
 \Delta Q_{Dp} = 4\Delta F_{Dp}^2 = \left\{
 \begin{array}{ll}
 2^{-p+2} & \mbox{for a fractional D$p$-brane} \\
 8 & \mbox{for the non-BPS D-particle} \;.
 \end{array}\right.
\ee
 Taking into account that $Q(0)=-1/16$ the absolute tension (and NSNS charge)
as a function of the background RR field $F$ is then given by \be
\label{tensionformula}
 Q(F) = 4F^2 - {1 \over 16} ~.
\ee

\subsection{String creation}

A physical picture for the tension and charge jumps
(and our original derivation of the above formula)
is provided by the string creation phenomenon.
This phenomenon is an example of the Hanany-Witten effect
\cite{HananyWitten}, in which two linked D-branes, {\em i.e.} two
D-branes which have a single mutually transverse direction, cross
each other \cite{StringCreation,BGL}. In order to conserve the
linking number, a string must be created between the two D-branes.

An example of a linked pair is the D0-D8 system in Type IIA. From
the D8-brane point of view, the linking number corresponds to a
world-volume electric charge induced by the D-particle, which flips
sign when the branes cross. Similarly, the D8-brane induces a charge
on the D-particle, which flips sign when the branes cross.
A unique feature of this particular linked brane system is that,
since the D-particle cannot support a net charge, the induced charge
must be cancelled by strings ending on it. For a single D8-brane
this requires an unusual $\pm 1/2$ string (depending on which side
of the D8-brane the D-particle sits), so the corresponding system is
presumably unphysical (a system with two D8-branes is more physical),
but the net change when the branes cross is the creation of a
single string.

The process of string creation can also be understood dynamically.
The ground state of the D0-D8 open string is a single massless
fermion, which leads to a repulsive one-loop effective potential
in the world-line theory of the D-particle,\footnote{The full
string effective potential vanishes by supersymmetry. From the
D-particle point of view there exists a tree-level term which
cancels the one-loop potential \cite{BGL}. The
former can be understood as coming from the (1/2) string attached
to the D-particle.}
\be
 V(x)=-|x|/2 \;.
\ee
This gives a jump in the force at $x=0$, which is exactly
compensated by the creation of a single string between the
D-particle and the D8-brane. More generally, the number of strings
which are created is given by the number of (massless) R ground
states of the open string between the two D-branes in the linked
system. This is equivalent to the supersymmetry index of the open
string, since the NS ground state is always massive in such a
system.

In our orbifold lines any two fractional D-branes are linked, and
therefore exhibit string creation. In particular, for a pair of
fractional D-particles on $\ofa$ there are $n=8$ massless R ground states,
and therefore 8 strings are created.
\begin{figure}
\centerline{\epsfxsize=110mm\epsfbox{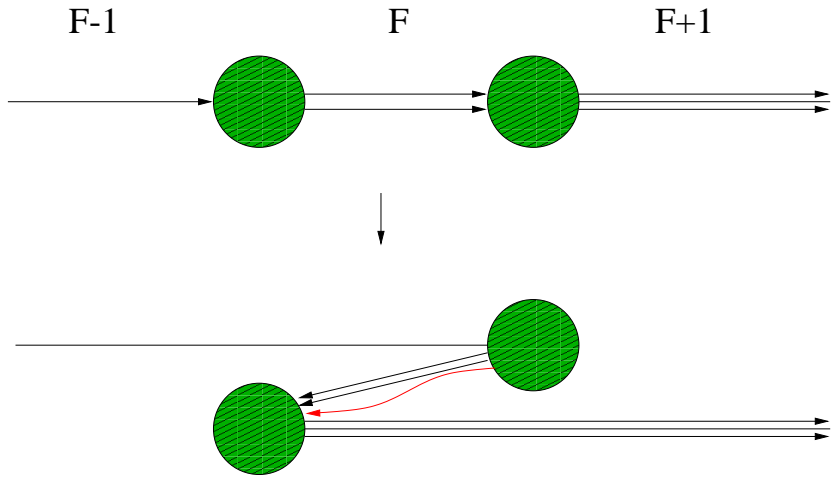}}
\medskip
\caption{String creation when two D0 branes are crossing on an
orbifold line. The wavy line is the newly created string.}
\label{fig1}
\end{figure}
We will now use this fact to compute the tension jump due to a single
D-particle. The two D-particles divide the orbifold line
into three regions (see figure \ref{fig1}): a semi-infinite line with $F-1$
field strength, a finite segment with $F$, and another semi-infinite
line with $F+1$. We denote the corresponding tensions
(or charges) $Q(F-1),Q(F)$ and $Q(F+1)$, respectively.
Next, we exchange the positions of the two D-particles,
and demand that $Q(F)$ remains unchanged in the middle segment,
\be
\label{recursive soln}
Q(F)=Q(F-1)-Q(F)+Q(F+1)-8 \;.
 \ee
The general solution to this recursion
equation is $Q(F)=4F^2 + a F + Q(0)$,
which is consistent with (\ref{tensionformula}).

To determine the value of the constant $a$ requires knowledge of
how many strings are attached to the D-particle to begin with,
and not just the number of strings which are created in the
exchange. As in the D0-D8 system, this is given by (minus) the
electric charge induced on a D-particle by a background field $F$.
By analogy with the D0-D8 system, we expect this charge to be of
the form $8\cdot F$. The situation here is somewhat more subtle,
in that the D-particle is both a source and a probe for the field
$F$. It is therefore not clear what should be taken as the
background value of $F$. A natural choice in the above case of the
two D-particles is the average of the two sides, $F+1/2$. The net
number of strings attached to the probe D-particle is then given
by $Q(F+1)-Q(F)=8 \cdot (F+1/2)$, which gives $Q(F)=4 F^2 + Q(0)$.

\subsection{The self intersection number}

Here we offer an alternative point of view for the computation of
$\Delta Q(F)$ which uses the self intersection in K-theory.  Type
IIA includes the following coupling\footnote{We thank E. Witten
for suggesting this. The normalization is inherited from the term
$1/6 \int C \wedge G \wedge G$ in 11d.}
 \be \label{charge from IIA action}
{1 \over 2} \int B_{NS} \wedge G^{(4)} \wedge G^{(4)}\;,
 \ee
where $G^{(4)}$ is the 4-form RR field strength, and all forms are
normalized to have integral fluxes. This term contributes
 $\Delta Q=(1/2) \int G^{(4)} \wedge G^{(4)}$ to the fundamental
string charge. We would like to perform this integral by using a
Poincar\'e duality to exchange $G^{(4)}$ with the dual 4-cycle, and
trade the wedge product with the intersection product. However,
since $\R^8/{\cal I}_8$ is singular, one needs to be careful.

The eight-dimensional orbifold $\R^8/{\cal I}_8$,
unlike $\R^4/{\cal I}_4$,
cannot be resolved in the usual manner. The necessary ''blow-up''
moduli, which appear as massless scalars in the twisted NSNS
sector of $\R^4/{\cal I}_4$, are absent for $\R^8/{\cal I}_8$.
The twisted NSNS sector of the orbifold line is purely massive.
This orbifold therefore corresponds to an isolated point in the
moduli space of Type II compactifications to two dimensions, {\em
i.e.} it is not a limit of Calabi-Yau four-folds.

Nevertheless, one expects it to correspond to a limit of some
smooth eight-dimensional manifold $X$, which is not Ricci-flat.
Since the only massless field in the twisted sector of this
orbifold for Type IIA ($\ofa$) is a RR vector,
we expect the above eight-dimensional manifold
to possess a single homology cycle $C$ of dimension four. The IIA
vector then corresponds to the reduction of $C^{(5)}$ on the
four-cycle.

Assuming that such a manifold $X$ exists, and that it has a 4
cycle $C$, we can take the dual of $G^{(4)}$ for the orbifold with
$F=1$ to be $C$. From (\ref{charge from IIA action}) we then find
$\Delta Q(F)=(1/2) (F \cdot C) \circ (F \cdot C)=(1/2)(C \circ C)
F^2$, from which we deduce
 \be |(C \circ C)|=n=8. \ee
 To be more precise one should replace the cohomological formula
 (\ref{charge
from IIA action}) by a K-theoretic one $\Delta Q(F)=(1/2)(F \cdot
y) \circ (F \cdot y)$, where $y$ is the class of the D-particle,
i.e. the generator of $K_{\Z_2,cpct}(\R^{8,0})$, and $y \circ y$
is a K-theory product.
It follows from (\ref{Kproduct}) that $y\circ y =8$.

We should point out, however, that so far we have not been able to
identify the manifold $X$. We found two manifolds $X$ satisfying
$\partial X=\RP{7}$, but neither one has the right properties.
Presumably these are not the right manifolds that can be
continuously deformed (continuous with respect to K-theory data)
into the singular $\C^4/\Z_2$. The first, $X_1$, is the disk bundle
${\cal O}(-2)$ over $\CP{3}$, which was discussed in
\cite{SethiO2}. It is gotten from the circle fibration ${\IS}^1
\to \RP{7} \to \CP{3}$, by filling the ${\IS}^1$. It does have a 4
cycle $C$, but it also has a redundant 2 cycle and a redundant 6
cycle. Moreover, the self intersection number of $C$ is $-2$.
Another space, $X_2$, can be gotten from the fibration ${\IS}^3 \to
{\IS}^7 \to {\IS}^4$, which induces $\RP{3} \to \RP{7} \to
{\IS}^4$. By filling the fiber with an Eguchi-Hanson space (EH)
one gets $\mbox{EH} \to X_2 \to {\IS}^4$. However, $X_2$ inherits
a redundant 2 cycle and a redundant 6 cycle from the 2 cycle of
EH. Thus neither $X_1$ nor $X_2$ meets our expectations.

\appendix
\section{An Orbifold K-theory Primer}

\subsection{Equivariant K-theory}

Consider a space $X$ acted on by a discrete symmetry group $G$. A
vector bundle $E$ over $X$, acted on by $G$
such that the projection $E\rightarrow X$ commutes with the
action of $G$, is called a {\em G-equivariant} bundle over $X/G$.
Pairs of such bundles $(E,F)$, modulo the usual identification
$(E\oplus H,F\oplus H)$ (where $H$ is also a $G$-equivariant
bundle), are classified in equivariant K-theory $K_G(X)$. These
groups satisfy many of the properties of ordinary K-theory, such
as Bott-periodicity, \be \label{Bott}
 K_G^{-k}(X)= K_G^{-k+2}(X) \;.
\ee
Two basic identities in equivariant K-theory which we will use
are
\be
 K_G(pt.) = R[G] \;,
\ee where $R[G]$ is the representation ring of $G$, and \be
\label{free-action}
 K_G(X)=K(X/G)
\ee if $G$ acts {\em freely} on $X$. As in ordinary K-theory, one
can also define the notion of {\em reduced} equivariant K-theory,
$\widetilde{K}_G(X)$, by requiring $E$ and $F$ to have equal rank,
or equivalently by removing the group evaluated at a ($G$-orbit of
a) generic point,
\be
 \widetilde{K}_G(X)=\mbox{ker}\left[K_G(X)\rightarrow
    K_G(\cup_i g_i ~pt.)\right] \qquad (g_i\in G)\;.
\ee Using (\ref{free-action}) we see that the latter group is
simply $K(pt.)=\Z$, so the usual decomposition holds, \be
\label{reducedK}
 K_G(X) = \widetilde{K}_G(X) \oplus \Z \;.
\ee

Let $X=\R^n$ and $G=\Z_2$, where the non-trivial element inverts
$l$ of the $n$ coordinates. The corresponding group is denoted
$K_{\Z_2}(\R^{l,m})$, where $m=n-l$. Since $\R^n$ is
(equivariantly)
contractable to a point, \be \label{Kpoint}
 K_{\Z_2}(\R^{l,m}) = R[\Z_2] = \Z \oplus \Z \;.
\ee
If the
bundles $E$ and $F$ are isomorphic at infinity, the pair $(E,F)$
is said to be {\em compactly supported}, and the classifying
groups are given by \cite{AtiyahSegal}
\be
\label{Kcompact}
 K_{\Z_2,cpct}(\R^{l,m})=
 K_{\Z_2}(\R^{l,m},{\IS}^{l,m}) = \left\{
 \begin{array}{ll}
  0 & \quad m\;\; \mbox{odd} \\
  \Z & \quad m\;\; \mbox{even},\; l\;\; \mbox{odd} \\
  \Z \oplus \Z & \quad m,l\;\; \mbox{even}\;,
 \end{array}\right.
\ee where ${\IS}^{l,m}$ denotes the $(n-1)$-sphere (at
infinity) in the space $\R^n/\Z_2$. There is no further reduction
of these groups, as the compact support condition already implies
that $E$ and $F$ have equal rank.
Higher K-theory groups are obtained by the suspension isomorphism,
\be
\label{suspension}
 K_{\Z_2,cpct}^{-k}(\R^{l,m}) = K_{\Z_2,cpct}(\R^{l,m+k})\;,
\ee
and are related by the usual Bott-periodicity (\ref{Bott}). In
particular this implies that
\be
\label{K-1point}
 K_{\Z_2}^{-1}(\R^{l,m})= K_{\Z_2}^{-1}(pt.) =
 K_{\Z_2,cpct}(\R^{0,1}) = 0 \;.
\ee

The computation of equivariant K-theory groups for
the spheres $X={\IS}^n$ is somewhat more involved in general.
However, for an $(n-1)$-sphere corresponding to the boundary of the
completely reflected space $\R^{n,0}$, denoted ${\IS}^{n,0}$,
the action of $\Z_2$ is free, and we can use (\ref{free-action}) to get
\cite{Adams}
\be
\label{adams}
 K_{\Z_2}({\IS}^{n,0}) = K(\RP{n-1}) =
  \Z \oplus \Z_{2^{[(n-1)/2]}} \;.
\ee
 The discrete torsion part corresponds to the reduced group
$\widetilde{K}(\RP{n-1})$, and can be determined as follows. The
generator of $\widetilde{K}(\RP{n-1})$ is $({\cal L}_{\C},{\bf
1})$, where ${\cal L}_{\C}$ is the complexification of the
canonical (real) line bundle ${\cal L}$ over $\RP{n-1}$. Recall
that a point $p$ in $\RP{n-1}$ corresponds to a real line through
the origin in $\R^n$. The fiber of ${\cal L}$ above the point $p$
is defined to be precisely this line. The line bundle can also be
expressed as $({\IS}^{n-1}\times\R)/\Z_2$, where the $\Z_2$
generator simultaneously reflects the sphere and the line. (For
example, the canonical line bundle over $\RP{1}\simeq {\IS}^1$
is the M\"obius strip.) One can, as usual, define local frames
over subsets $U_i$ of $\RP{n-1}$ which trivialize the bundle
locally, but the $\Z_2$ twist prevents these frames from
combining into a global non-degenerate frame. Such a frame can
however be constructed by considering multiple copies of the
bundle, and is given by $\Gamma^i\cdot \widehat{x}^i$, where
$i=1,\ldots,n$. This requires $2^{[(n-1)/2]}$ copies of ${\cal
L}_{\C}$, since that is the dimension of the irreducible complex
spinor representation of $SO(n)$. (Note that for $n=2$, ${\cal
L}_{\C}$ is twice the real M\"obius bundle, and is therefore
trivial.)

The different groups are part of a long exact sequence given by
\begin{eqnarray}
\label{longsequance}
 \cdots \rightarrow K_{\Z_2}^{-1}(\R^{l,m}) & \rightarrow &
 K_{\Z_2}^{-1}({\IS}^{l,m}) \rightarrow
 K_{\Z_2,cpct}(\R^{l,m}) \nonumber \\
 & \stackrel{i}{\rightarrow} & \widetilde{K}_{\Z_2}(\R^{l,m})
 \rightarrow \widetilde{K}_{\Z_2}({\IS}^{l,m}) \rightarrow
 K_{\Z_2,cpct}^{-1}(\R^{l,m}) \rightarrow \cdots \;,
\end{eqnarray}
which, for $l=n$ and $m=0$ reduces to
\be
    0  \rightarrow K_{\Z_2}^{-1}({\IS}^{n,0})  \rightarrow  \Z'\oplus \Z
  \stackrel{i}{\rightarrow} \Z
  \rightarrow \Z_{2^{[(n-1)/2]}} \rightarrow 0 \;.
\ee
It follows that $K_{\Z_2}^{-1}({\IS}^{n,0})=\Z'$, and that
the map $i$ is a multiplication by $2^{[(n-1)/2]}$.

\subsection{Atiyah-Hirzebruch spectral sequence}

There is a useful technique of successive approximations to
K-theory known as the Atiyah-Hirzebruch spectral sequence (AHSS).
We start by triangulating the $n$-dimensional space $X$ with
successive trees $X^p$, where $p=0,\ldots,n$, and forming the
filtration \be
 K_n(X)\subseteq K_{n-1}(X) \subseteq\cdots\subseteq K_0(X)=K(X)\;,
\ee
where $K_p(X)$ is defined to contain classes which are trivial on the
$(p-1)$-tree $X^{p-1}$. For example $K_1(X)=\widetilde{K}(X)$.
Analogous filtrations can be formed for $K^{-1}(X)$, and for equivariant
K-theory. The second step is to define a graded complex associated
to the K-theory group in question, {\em e.g.}
\be
 \mbox{Gr}K^{-k}(X) = \oplus_p K^{-k}_p(X)/K^{-k}_{p+1}(X) \;.
\ee
The successive ratios forming the graded complex are computed
using a spectral sequence of groups $E_r^{p,q}$ with differential maps
$d_r: E_r^{p,q} \rightarrow E_r^{p+r,q-r+1}$,
such that each $E_{r+1}^{p,q}$ is
the cohomology of $d_r$. The first terms in the sequence $E_1^{p,q}$
are given by singular $p$-cochains on $X$ with values in $K^{-q}(pt.)$,
\be
 E_1^{p,q} = C^p(X;K^{-q}(pt.)) = \left\{
 \begin{array}{ll}
 C^p(X;\Z) & \quad q\;\; \mbox{even} \\
 0 & \quad q\;\; \mbox{odd}\;.
 \end{array}\right.
\ee
The differential $d_1$ can be identified with the simplicial co-boundary
operator acting on these co-chains. The second set of terms in the
sequence is therefore given by
\begin{equation}
\label{E2}
E_2^{p,q} = H^p(X,K^{-q}(pt.)) = \left\{
 \begin{array}{ll}
 H^p(X;\Z) & \quad q\;\; \mbox{even} \\
 0 & \quad q\;\; \mbox{odd}\;.
 \end{array}\right.
\end{equation}
The successive ratios are given by the limit of this sequence,
\begin{equation}
\label{ratios}
 K^{-k}_p(X)/K^{-k}_{p+1}(X) = E_{\infty}^{p,-(p+k)}\;,
\end{equation}
although in the cases we are interested in the sequence terminates
at $E_2^{p,q}$.
For example, for ${\IS}^{8,0}$ we get
\be
\begin{array}{lclcl}
 \mbox{Gr}K_{\Z_2}({\IS}^{8,0}) &=& H^{even}(\RP{7};\Z)
 &=& \Z \oplus \Z_2\oplus\Z_2\oplus\Z_2 \\
 \mbox{Gr}K^{-1}_{\Z_2}({\IS}^{8,0}) &=& H^{odd}(\RP{7};\Z)
 &=& \Z \;.
\end{array}
\ee
Once we have the graded complex, computing the filter groups
$K_p^{-k}$, and in particular the K-theory group itself, requires a
solution of extension problems of the form
\be
\label{extension}
 0\rightarrow K_{p+1}^{-k}(X) \rightarrow
 K_{p}^{-k}(X) \rightarrow
 K_{p}^{-k}(X)/K_{p+1}^{-k}(X) \rightarrow 0 \;.
\ee
In the second case the solution is unique, and gives
\be
 K^{-1}_{\Z_2}({\IS}^{8,0}) = \Z \;,
\ee
but in the first case there are three possible solutions:
$\Z\oplus \Z_2^3$, $\Z\oplus \Z_2\oplus\Z_4$, and
$\Z\oplus \Z_8$. From the previous
discussion we know that the correct answer is
$K_{\Z_2}({\IS}^{8,0}) = \Z\oplus \Z_8$.

Although the extension problem precludes the AHSS as a useful way of
computing the latter group, it does offer some physical insight
into the nature of the discrete torsion ``enhancement'' in K-theory
compared to cohomology. The point is that the filter groups $K_p$
($p\geq 1$) can also be thought of as K-theory groups which classify
RR fields starting with a D$(9-p)$-$\overline{\mbox{D}(9-p)}$-brane pair.
The filter groups for $K_{\Z_2}({\IS}^{8,0})$ are therefore given by
\be
 K_{p}({\IS}^{8,0}) = \widetilde{K}_{\Z_2}({\IS}^{9-p,0})\;.
\ee
Using (\ref{adams}) we get
\begin{eqnarray}
 K_1&=&K_2=\Z_8 \nonumber\\
 K_3&=&K_4=\Z_4 \nonumber\\
 K_5&=&K_6=\Z_2  = H^6(\RP{7};\Z) \nonumber\\
 K_7&=&0\;,
\end{eqnarray}
where the identification of the $\Z_2$ above with $H^6$
follows from (\ref{E2}), (\ref{ratios}), and the fact that $K_7=0$.
Consider now the two sequences of the form (\ref{extension}) with
$p=4$ and $p=2$,
\be
\begin{array}{ccccccccc}
0 & \rightarrow & K_5 & \rightarrow & K_4 & \rightarrow &
  H^4(\RP{7};\Z) & \rightarrow & 0 \\
  & & \parallel & & \parallel & & \parallel & & \\
   & & \Z_2 & \rightarrow & \Z_4 & \rightarrow & \Z_2 && \\[10pt]
0 & \rightarrow & K_3 & \rightarrow & K_2 & \rightarrow &
  H^2(\RP{7};\Z) & \rightarrow & 0 \\
  & & \parallel & & \parallel & & \parallel & & \\
   & & \Z_4 & \rightarrow & \Z_8 & \rightarrow & \Z_2 &&
\end{array}
\ee
Denote the generator of $\Z_8$ by $x$, the generator of $\Z_4$ by $y$,
and the generators of the three even ($\Z_2$) cohomologies
$x^{(2)}, x^{(4)}$, and $x^{(6)}$. It then follows from the above sequences
that
\be
 x=x^{(2)}\;,\; 2x=x^{(4)}\;,\; 4x=x^{(6)}\;,
\ee
which gives equation (\ref{Ktorsion}).

\subsection{Hopkins' groups $K_{\pm}$}

If the generator of the $\Z_2$ includes the operator $(-1)^{F_L}$
in addition to inversion, the bundles are not equivariant, but
rather we are given an isomorphism $\lambda : (E,F)\rightarrow
(g^*F,g^*E)$ with $(\lambda g^*)^2=1$. The corresponding groups
are denoted $K_{\pm}(X)$. These groups have not been studied as
extensively as equivariant K-theory, so far less is known about
them (see \cite{Gukov} for related results).

A theorem due to Hopkins (unpublished) asserts that
\be
\label{hopkins}
 K_{\pm,cpct}(X) = K_{\Z_2,cpct}(X\times\R^{1,1}) \;.
\ee
This relation can be understood as the failure of ordinary Bott
periodicity $K_{cpct}(X)=K_{cpct}(X\times\R^2)$, due to the
orientation reversal of the $\R^2$.\footnote{This was pointed out to us
by M. Atiyah.}
It should be contrasted with
the case in which the orientation is preserved,
\be
 K_{\Z_2,cpct}(X) = K_{\Z_2,cpct}(X\times\R^{0,2}) \;.
\ee
In particular this implies
\be
\begin{array}{lclclcc}
 K_{\pm}(\R^{l,m}) & = & K_{\pm}(pt.) &=& K_{\pm,cpct}(\R^{0,0}) &=& 0 \\
 K_{\pm}^{-1}(\R^{l,m}) &=& K_{\pm}^{-1}(pt.) &=&
 K_{\pm,cpct}^{-1}(\R^{0,0}) & = &\Z \;,
\end{array}
\ee
where in the last step we used the suspension isomorphism
(\ref{suspension}), which continues to hold for $K_{\pm}$.

For the spheres, one can again try the AHSS.
The analysis is very similar to the one above for
equivariant K-theory, except that the singular cochains and
cohomologies valued in $\Z$ are replaced by those valued
in the twisted $\widetilde{\Z}$. This is because all RR fields
are odd under $(-1)^{F_L}$. The corresponding graded complexes
are therefore given by
\be
\begin{array}{lclcl}
 \mbox{Gr}K_{\pm}({\IS}^{8,0}) &=& H^{even}(\RP{7};\widetilde{\Z})
 &=& 0 \\
 \mbox{Gr}K^{-1}_{\pm}({\IS}^{8,0}) &=& H^{odd}(\RP{7};\widetilde{\Z})
 &=& \Z_2 \oplus \Z_2 \oplus \Z_2 \oplus \Z_2 \;.
\end{array}
\ee
In particular, this implies that $K_{\pm}({\IS}^{8,0})=0$.
The second group $K^{-1}_{\pm}({\IS}^{8,0})$ suffers from the
same extension
problems as $K_{\Z_2}({\IS}^{8,0})$, and it is reasonable to guess
that a complete enhancement occurs to $\Z_{16}$. From
(\ref{hopkins}) it follows that
\be
 K_{\pm}^{-1}({\IS}^{n,0}) = K_{\Z_2,cpct}^{-1}({\IS}^{n,0}\times\R^{1,1})
 = K_{\Z_2,cpct}({\IS}^{n,0}\times\R^{1,0}) \;,
\ee
where in the second equality we have used the suspension isomorphism
and Bott periodicity.
The action on the space $X={\IS}^{n,0}\times\R^{1,0}$ is free,
and gives $X/\Z_2 \sim {\cal L}_{n-1}$, where ${\cal L}_{n-1}$ is the
canonical real line bundle over $\RP{n-1}$.
Since the compactification of ${\cal L}_{n-1}$ is topologically
the same as $\RP{n}$ we find
\be
 K_{\pm}^{-1}({\IS}^{n,0}) = \widetilde{K}(\RP{n})
 = \Z_{2^{[n/2]}} \;,
\ee
and in particular $K_{\pm}^{-1}({\IS}^{8,0}) = \Z_{16}$.
Similarly, one can show that $K_{\pm}({\IS}^{n,0})=0$.\footnote{We thank
M. Atiyah for discussions on these issues.}

\vspace{0.5cm} \noindent {\bf Acknowledgments}

\noindent We would like to thank E. Witten, M. Atiyah and S. Sethi
for useful conversations. B.K. would like to thank I. Brunner, A.
Hanany, A. Schwarz, and members of the IAS group, especially K.
Dasgupta, E. Diaconescu, J. Maldacena and  N. Seiberg for
discussions. B.K. also thanks the Hebrew University in Jerusalem
for hospitality during the course of the work. E.G. would like to
thank the ITP at Santa Barbara which is supported in part by the
National Science Foundation under Grant No. PHY99-07949 for their
hospitality while this work was being completed.

O.B. was supported in part by the DOE under grant no.
DE-FG03-92-ER40701, and by a Sherman Fairchild Prize Fellowship.
The work of E.G. was supported by NSF grant no. 0070928 and by
Frank and Peggy Taplin.  The work of B.K. was supported by DOE
grant no. DE-FG02-90ER4052 and by a Raymond and Beverley Sackler
Fellowship.

\end{document}